\documentclass[12pt,final,english]{iopart}
\usepackage[T1]{fontenc}
\usepackage[latin1]{inputenc}
\usepackage{graphicx}
\usepackage{setspace}
\usepackage[usenames,dvipsnames]{color}
\makeatletter

\newcommand{\TAE}{TAE}
\newcommand{\TAEs}{TAEs}
\newcommand{\MHD}{MHD}

\makeatother
\begin{document}

\title
{A singular finite element technique for calculating continuum damping of Alfv\'{e}n eigenmodes}

\author{G.W. Bowden}
\address{Research School of Physical Sciences and Engineering, Australian National University, Acton 0200, ACT Australia}
\author{M.J.~Hole}
\address{Research School of Physical Sciences and Engineering, Australian National University, Acton 0200, ACT Australia}

\begin{abstract}
Damping due to continuum resonances can be calculated using dissipation-less ideal magnetohydrodynamics (\MHD ) provided that the poles due to these resonances are properly treated. We describe a singular finite element technique for calculating the continuum damping of Alfv\'{e}n waves. A Frobenius expansion is used to determine appropriate finite element basis functions on an inner region surrounding a pole due to the continuum resonance. The location of the pole due to the continuum resonance and mode frequency are calculated iteratively using a Galerkin method. This method is used to find the complex frequency and mode structure of a toroidicity-induced Alfv\'{e}n eigenmode (\TAE ) in a large aspect ratio circular tokamak and are shown to agree closely with a complex contour technique.

\end{abstract}


\maketitle

\section{Introduction}
A variety of lightly damped Alfv\'{e}nic modes can occur in magnetically confined plasmas. Interaction with fast particles can destabilise these modes, leading to a loss of confinement of these particles. In fusion applications, such losses could reduce heating and damage plasma facing components. One such mode is the toroidicity-induced shear Alfv\'{e}n eigenmode (\TAE), which occurs due to the coupling of successive poloidal harmonics in a toroidal plasma. This coupling results in frequency gaps in the spectrum of spatially-localised modes resonant on particular flux surfaces. \TAEs\, which have extended radial structure, can exist in these frequency gaps. One potentially significant source of damping for these global modes is the resonant transfer of energy from these modes to the strongly damped continuum modes.

Physically, continuum damping represents dissipative effects, such as charge separation and mode conversion which occur near continuum resonances. Such behaviour could be described using two-fluid and kinetic plasma models. However, it is not necessary to model these mechanisms to compute continuum damping. Continuum damping can be calculated using resistive \MHD\ as the limit of damping as resistivity is reduced to zero \cite{damping_of_GAEs_in_tokamaks_due_to_resonant_absorption}. In ideal \MHD\, a continuum resonance corresponds to a regular singular point. The correct treatment of such a singularity is dictated by the causality condition, analogous to the analysis of Landau damping of plasma oscillations \cite{landau_damping}. Analytic \cite{continuum_damping_of_high_n_TAWs,continuum_damping_of_ideal_TAEs,resonant_damping_of_TAEs_in_tokamaks,continuum_damping_of_low_n_TAEs} and numerical \cite{computational_approach_to_continuum_damping_in_3D_published}  methods have been developed for calculating continuum damping using ideal \MHD\ based on this condition. In this paper we outline a numerical method in which singular finite elements are used to ensure that the singularity in the ideal \MHD\ treatment of continuum damping is properly represented. While we describe the calculation of continuum damping of a \TAE\, the method used could be adapted to the continuum damping of other global modes in toroidal plasmas.

Singular finite elements are used in a number of fields, with examples of applications occurring in fracture mechanics \cite{sfe_for_crack_propagation}, electromagnetism \cite{edge_elements_for_eddy_current_problems} and viscous flow \cite{sfe_for_Stokes_flow}. To our knowledge, singular finite element methods have only been employed in one case in plasma physics to solve the ideal \MHD\ Newcomb equation describing a marginally stable cylindrical plasma \MHD\ \cite{galerkin_method_for_DEs_with_regular_singular_points}. This method was applied to the analysis of resistive instabilities. In these cases the asymptotic behaviour around and location of the singular point is known. Singular finite elements have not previously been applied to \MHD\ problems with finite frequency oscillations. In the case we analyse, the location of continuum resonances is dependent on the \TAE\ eigenvalue. Consequently, determining this eigenvalue using the singular finite element method is an iterative process, in which the estimated location of the resonance must be updated after each iteration.

In section \ref{sec:TAE_equation} of this paper we outline a \TAE\ model due to Berk \textit{et al.}\ and show that the resulting eigenvalue equation can be expressed in terms of Hermitian operators. The Frobenius method is applied to this expression to determine the form of the continuum resonance singularities in section \ref{sec:singularities}. Subsequently, in section \ref{sec:FEM} a finite element method is described which incorporates elements with this form. This method is applied to a \TAE\ in section \ref{sec:verification} and verified by comparison to the complex contour method.

\section{Shear Alfv\'{e}n eigenmode equation} \label{sec:TAE_equation}
In this paper we consider a \TAE\ in a low $\beta$, large aspect ratio circular tokamak with a perfectly conducting wall at the edge of the plasma. For this case Berk \textit{et al.}\ have derived the following coupled mode equation for shear Alfv\'{e}n waves in ideal \MHD \cite{continuum_damping_of_low_n_TAEs}:
\begin{eqnarray} 
\noindent \frac{d}{dr}\left [r^3\left (\frac{\omega^2}{{v_A}^2}-k_{\parallel m}^2 \right ) \frac{dE_m}{dr}\right]+\frac{d}{dr}\left (\frac{\omega^2}{{v_A}^2} \right )r^2E_m-\left (m^2-1 \right )\left(\frac{\omega^2}{v_A^2}-k_{\parallel m}^2 \right )rE_m && \nonumber \\
\noindent +\frac{d}{dr}\left [\frac{5\epsilon r^4}{2a} \left (\frac{dE_{m+1}}{dr}+\frac{dE_{m-1}}{dr} \right ) \right ]=0 , \label{eq:wave_equation}
\end{eqnarray}
in which $E_m$ is the $m$'th poloidal Fourier component of $\frac{\delta \Phi}{r}$, where $\delta \Phi$ is the perturbation to the electric potential due to the wave. The gauge is set such that the magnetic vector potential $\mathbf{A}$ associated with these oscillations is parallel to the equilibrium magnetic field. The variable $r$ is the radial coordinate in the flux-type straight-field-line coordinates defined by Berk \textit{et al.}\ \cite{continuum_damping_of_low_n_TAEs}. The inverse aspect ratio $\epsilon$, is the ratio of the minor radius $a$ to the major radius $R_0$, $\omega$ is the angular frequency of the mode and $k_{\parallel}=\frac{1}{R_0}\left (n - \frac{m}{q} \right )$ is the wave number parallel to the magnetic field for poloidal and toroidal mode number $m$ and $n$ respectively. For a particular flux surface $v_A = \frac{B}{\mu_0 \rho}$ is the Alfv\'{e}n speed and $q$ is the safety factor.

Dividing by $a$, equation (\ref{eq:wave_equation}) can be written more compactly as:
\begin{eqnarray} 
\Omega^2 L_{\Omega} \left [ E_j \right ] - L_{k} \left [ E_j \right ] = \frac{d}{dx} \left [  \left ( \Omega^2 D_{\Omega,i,j} \left ( x \right ) - D_{k,i,j} \left ( x \right ) \right ) \frac{dE_j}{dx} \right ] && \nonumber \\
+ \left ( \Omega^2 A_{\Omega,i,j} \left ( x \right ) - A_{k,i,j} \left ( x \right ) \right )E_j = 0 . \label{eq:matrix_equation}
\end{eqnarray}
Here we define:
\begin{eqnarray}
D_{\Omega,i,j} = \frac{N}{N_0} \left ( x^3 \delta_{i,j} + \frac{5}{2} \epsilon x^4 \left ( \delta_{i-1,j} \delta_{i+1,j} \right ) \right ) ,  && \\
D_{k,i,j} = x^3 \left (n - \frac{m}{q} \right ) ^2 \delta _{i,j} , && \\
A_{\Omega,i,j} = \left ( x^2 \frac{d}{dx} \left (\frac{N}{N_0} \right ) - \left (m^2 - 1 \right ) x \left (\frac{N}{N_0} \right ) \right ) \delta_{i,j} , && \\
A_{k,i,j} = -\left ( m^2 - 1 \right ) \left (n - \frac{m}{q} \right ) ^2 x \delta_{i,j} ,
\end{eqnarray}
in which $x = \frac{r}{a}$ is the normalised radial coordinate and $\Omega = \frac{\omega R_0}{v_{A,0}}$ is the normalised frequency, where $v_{A,0}$ is the Alfv\'{e}n speed at the magnetic axis. The operators $L_{\Omega}$ and $L_{k}$ can be shown to be Hermitian. As ideal \MHD\ is non-dissipative, the corresponding operators must be Hermitian for arbitrary geometry. Thus the equations of linearised ideal \MHD\ spectral analysis may always be expressed in the form seen in equation (\ref{eq:matrix_equation}), regardless of geometry and simplifying assumptions. Therefore, the methods described here for a simplified case could in principle be applied in three-dimensional geometry or where there is interaction between shear-Alfv\'{e}n and magneto-sonic waves.

If operator $L$ is Hermitian, then:
\begin{equation}
\int_{a}^{b} \left ( L \left [ \mathbf{y} \right ] \right )^{\dagger} {\mathbf{y}}' dx = \int_{a}^{b} \mathbf{y}^{\dagger} \left ( L \left [ {\mathbf{y}}' \right ] \right ) dx , \label{eq:hermitian}
\end{equation}
for all $\mathbf{y}$ and ${\mathbf{y}}'$. If $L$ is a second order differential operator:
\begin{equation}
L \left [ \mathbf{y} \right ] = \left ( L_2 \left (x \right ) \frac{d^2}{dx^2} + L_1 \left (x \right ) \frac{d}{dx} + L_0 \left (x \right ) \right )\left [ \mathbf{y} \right ] . \label{eq:sodo}
\end{equation}
Substitute equation (\ref{eq:sodo}) into equation (\ref{eq:hermitian}) and integrate by parts,
\begin{eqnarray}
\int_{a}^{b} \left ( L_2 \left (x \right ) \frac{d^2 \mathbf{y}}{dx^2} + L_1 \left (x \right ) \frac{d \mathbf{y}}{dx} + L_0 \left (x \right ) \mathbf{y} \right )^{\dagger} {\mathbf{y}}' dx && \nonumber \\
%= \int_{a}^{b} \left ( \frac{d^2 {\mathbf{y}}^{\dagger}}{dx^2} L_2 \left (x \right ) {\mathbf{y}}' + \frac{d {\mathbf{y}}^{\dagger}}{dx} \left ( \frac{d L_2 \left (x \right ) }{dx} - L_1 \left (x \right ) \right ) {\mathbf{y}}' + {\mathbf{y}}^{\dagger} L_0 \left (x \right ) {\mathbf{y}}' \right ) dx && \nonumber \\
%+ \left. \left \[ {\mathbf{y}}^{\dagger} L_2 \left (x \right ) \frac{d{\mathbf{y}}'}{dx} - \frac{d {\mathbf{y}}^{\dagger}}{dx}  L_2 \left (x \right ) {\mathbf{y}}' + {\mathbf{y}}^{\dagger} \left ( L_1 \left (x \right ) - \frac{d L_2 \left (x \right )}{dx} \right ) {\mathbf{y}}' \right \] \right | ^b_a && \nonumber \\
= \int_{a}^{b} \left ( \frac{d^2 {\mathbf{y}}^{\dagger}}{dx^2} L_2 \left (x \right ) {\mathbf{y}}' + \frac{d {\mathbf{y}}^{\dagger}}{dx} \left ( \frac{d L_2 \left (x \right ) }{dx} - L_1 \left (x \right ) \right ) {\mathbf{y}}' + {\mathbf{y}}^{\dagger} L_0 \left (x \right ) {\mathbf{y}}' \right ) dx .
\end{eqnarray}
The last step assumes that $\mathbf{y} = 0$ for $x = a$ and $x = b$. Hence, to ensure that $L$ is Hermitian, we require that:
\begin{eqnarray}
L_2 = {L_2}^{\dagger} , && \\
2 \frac{d L_2}{dx} - L_1 = {L_1}^{\dagger} , && \\
\frac{d^2 L_2}{dx^2} - \frac{d L_1}{dx} + L_1 = {L_0}^{\dagger} .
\end{eqnarray}
Therefore, for Hermitian matrices $L_1$ and $L_0$ this implies that $\frac{d L_2}{dx} = L_1 $, and thus the equation $L E_m = 0$ has the same form as equation (\ref{eq:matrix_equation}). Clearly, $L_1$ and $L_0$ are Hermitian for a circular tokamak with the large aspect ratio approximation described in equation (\ref{eq:matrix_equation}). However, it can also be shown that the force operator for linearised ideal \MHD\ can be expressed in a symmetric form \cite{Principles_of_MHD}. Therefore, it follows that matrices $L_1$ and $L_0$ will be Hermitian for arbitrary geometry. Thus the equations of ideal \MHD\ spectral analysis may always be expressed in the form seen in equation (\ref{eq:matrix_equation}), regardless of geometry and simplifying assumptions. Therefore, the methods described here for a simplified case could in principle be applied in three-dimensional geometry or where there is interaction between shear-Alfv\'{e}n and magneto-sonic waves.

\section{Continuum resonance singularities} \label{sec:singularities}
Regular singularities occur for $x$ such that the determinant of the matrix $D_{i,j} = \left ( \Omega^2 D_{\Omega,i,j} \left ( r \right ) - D_{k,i,j} \left ( r \right ) \right )$ is zero. At this point the inverse of the operator $L = \Omega ^2 L_{\Omega}-L_{k}$ is unbounded. The behaviour of the wave function near this pole can be found using a Frobenius expansion. Let $z = x - x_r$, where $x_r$ is the location of a pole due to a continuum resonance, and let $A_{i,j} = \left ( \Omega^2 A_{\Omega,i,j} \left ( r \right ) - A_{k,i,j} \left ( r \right ) \right )$. In general $x_r$ will be displaced into the complex plane as $\Omega$ become complex due to continuum damping. Where the inverse of $D$ exists, from equation (\ref{eq:matrix_equation}) we can derive the following expression:
\begin{equation}
\frac{d^2 E_j}{dz^2} + D_{i,j}^{-1} \frac{d D_{i,j}}{dz} \frac{dE_j}{dz} + D_{i,j}^{-1} A_{i,j} E_j = 0 . \label{eq:simplified}
\end{equation}
Assume that $\left. \frac{d \left \| D_{i,j} \right \|}{dx} \right |_{x=x_r} \neq 0$. That is, consider a case where the continuum resonance does not coincide with a stationary point of the continuous spectrum. Hence, near the pole, the inverse of matrix $D_{i,j}$ can be approximated in terms of its adjugate as:
\begin{equation}
D_{i,j}^{-1} \approx \frac{\textup{adj} \left ( D_{i,j} \right )}{z \left. \frac{d \left \| D_{i,j} \right \|}{dx} \right |_{x=x_r}} .
\end{equation}
Thus it is possible to write equation (\ref{eq:simplified}) as
\begin{equation}
\frac{d^2 E_j}{dz^2} + \frac{1}{z} M_{i,j} \frac{dE_j}{dz} + \frac{1}{z} N_{i,j} E_j = 0 , \label{eq:Fobenius}
\end{equation}
where
\begin{eqnarray}
M_{i,j} = \left [ \left. \frac{\textup{adj} \left ( D \right ) _{i,k}}{\frac{d \left \| D_{i,j} \right \|}{dx} } \frac{d D_{k,j}}{dx}  \right ] \right |_{x=x_r} = \delta_{i,j}, && \\ 
N_{i,j} = \left [ \left. \frac{\textup{adj} \left ( D \right ) _{i,k}}{\frac{d \left \| D_{i,j} \right \|}{dx} }  A_{k,j}  \right ] \right |_{x=x_r} .
\end{eqnarray}
Using the Frobenius method, express the solution near the resonance as:
\begin{equation}
E_j = z^k \sum_{l=0}^{\infty } a_{l,j} z^l ,
\end{equation}
where $a_{0,i} \ne 0$ for some $i$. Combining this requirement with equation (\ref{eq:Fobenius}) leads to the indicial equation:
\begin{equation}
k^2 a_{0,i} = 0 .
\end{equation}
Hence, the indicial equation has the double root $k = 0$. Therefore, the indicial equation does not provide two linearly independent solutions. As a consequence, the solution will have general form:
\begin{equation}
E_j = z^k \sum_{l=0}^{\infty } \left ( a_{l,j} z^l + b_{l,j} z^l \ln \left ( z \right )\right ) ,
\end{equation}
where $a_{0,i} \ne 0$ or $b_{0,i} \ne 0$ for some $i$. Thus, the indicial equations are:
\begin{eqnarray}
k^2 a_{0,i} + 2k b_{0,i} = 0 , && \\
k^2 b_{0,i} = 0 .
\end{eqnarray}
For these equations $k = 0$ remains a solution, and if $b_{0,i} = 0$ the previous solution is recovered. Additional non-trivial solutions are found with $b_{0,i} \ne 0$ for some $i$. Thus the solution in the vicinity of the resonance can be approximated to first order as $E_m = a_m + b_m \ln \left ( x - x_r \right )$.

It is possible to express the normalised real frequency $\Omega_r$ and damping $\Omega_i$ as normalised complex frequency $\Omega = \Omega_r + i \Omega_i$. For complex $\Omega$, the pole due to the continuum resonance will generally have an imaginary component. Consequently, it is necessary to define an analytic continuation of the logarithmic function in the vicinity of the complex pole. The causality condition requires that the logarithmic function is found by analytic continuation of the function on a path that with the real axis encloses the singularity. Physically, this derives from the requirement that a perturbation to the plasma precedes the response it causes. A branch cut exists where $\Re \left (x \right ) = \Re \left (x_r \right )$ and $\Im \left (x \right ) < \Im \left (x_r \right )$ if $\Im \left (x_r \right ) > 0$ or $\Im \left (x \right ) > \Im \left (x_r \right )$ if $\Im \left (x_r \right ) < 0$. Thus the logarithmic function in the Frobenius approximation will be:
\begin{equation}
\ln _{\pm} \left ( x - x_r \right ) = \ln \left | x - x_r \right | \mp  \pi i \arg \left ( x - x_r \right ) \pm 2 \pi i \Theta \left ( \Re \left ( x - x_r \right ) \right ) , \label{eq:aclog}
\end{equation}
where $\textup{sgn} \left ( \Im  \left ( x_r \right ) \right ) = \pm 1$ and $\Theta \left ( x_r \right )$ is the Heaviside step function.

\section{Finite element method} \label{sec:FEM}
In the Galerkin method the eigenvalue problem $\left ( L_a - \lambda L_b \right ) \mathbf{u} = 0$ is discretised by deriving a weak formulation of the problem and approximating the solution as a linear combination of a finite number of basis functions. The weak formulation is expressed in terms of bi-linear forms $a\left ( u, v\right )$ and $b\left ( u, v\right )$ defined for $u,v \in V$ where $V$ is a Hilbert space. The eigenfunction $u \in V$ and eigenvalue $\lambda$ are such that $a\left ( u, v\right ) - \lambda b\left ( u, v\right ) = 0$ $\forall v \in V$. This problem is discretised by solving for $u^h , v^h \in V^h$ where $V^h$ is a $h$-dimensional subset of $V$, representing the space spanned by a finite set of basis functions. This allows the problem to be represented as a generalised matrix eigenvalue problem, for which efficient numerical solution procedures exist. The solution obtained using this method is such that its error $e^h$ (the difference between it and the exact solution to the original eigenvalue problem) is Galerkin orthogonal to the space spanned by the chosen basis functions. That is $a\left ( e^h , v^h \right ) - \lambda b\left ( e^h, v^h \right ) = 0$ $\forall$ $v^h$. Thus this solution represents a projection of the exact solution onto the chosen space $V^h$. The accuracy of the solution obtained depends on how accurately the eigenfunction can be approximated by a linear combination of the chosen basis elements. Finite element method results can be made to converge with an increasing number of basis functions.

For simplicity, triangular functions with uniform spacing were chosen to form the basis set. These are defined as follows,
\begin{equation}
v_{m,i,j} \left ( x \right ) = \left \{ \begin{array}{rcl}
\left ( 1 - \frac{\left |  x - x_i \right |}{\Delta} \right ) \delta_{m,j} &\mbox{ for } x \in \left (x_i - \Delta , x_i + \Delta \right ) \\
0 &\mbox{for } x \not\in \left (x_i - \Delta , x_i + \Delta \right )         
\end{array} \right.
\end{equation}
%\begin{equation}
%v_{m,i,j} \left ( x \right ) = \begin{cases} \left ( 1 - \frac{\left |  x - x_i \right |}{\Delta} \right ) \delta_{m,j} &\mbox{ for } x \in \left (x_i - \Delta , x_i + \Delta \right )
%\\ 
%0 &\mbox{for } x \not\in \left (x_i - \Delta , x_i + \Delta \right )
%\end{cases},
%\end{equation}
where the centre of the $i$th basis function is located at $x_i = \frac{i-1}{N-1}$, $N$ is the number of elements and $\Delta = \frac{1}{N-1}$ is the spacing between the centres of adjacent functions. This gives a piecewise linear approximate solution $E_m = \sum_{i=0}^{N} \phi _{m,i,j} v_{m,i,j} \left ( x \right )$.

It is possible to express the \TAE\ wave equation in terms of a sesquilinear form by taking the scalar product with a function $E_m'$ and integrating from $x = 0$ to $x = 1$. In the absence of continuum resonances this leads to the expression:
\begin{equation}
\Omega^2 \int_{0}^{1} \frac{dE_i^*}{dx} \left [  \left (D_{\Omega,i,j} \left ( x \right ) - A_{\Omega,i,j} \left ( x \right ) \right )  \right ] \frac{dE_j}{dx} dx = \int_{0}^{1} E_i^* \left [ D_{k,i,j} \left ( x \right ) - A_{k,i,j} \left ( x \right ) \right ] E_j dx
\end{equation},
for all continuous $E_m'$ where $E_m\left ( 0 \right ) = E_m'\left ( 0 \right ) = E_m\left ( 1 \right ) = E_m'\left ( 1 \right ) = 0$. However, due to the discontinuity associated with a logarithmic singularity, it is necessary to exclude the continuum resonance from the integration. Let $x_r^-$ and $x_r^+$ be the real valued lower and upper bounds of a region containing the continuum resonance $\Re \left ( x_r \right )$, where $\Re \left ( x_r \right ) - x_r^- \ll 1 $ and $x_r^+ - \Re \left ( x_r \right ) \ll 1 $. If the region where $x \in \left [ x_r^-, x_r^+ \right ]$ is removed from the integration, this leads to the appearance of surface terms:
\begin{eqnarray}
\hspace{-1 cm} \noindent \Omega^2 \left [\left. E_i^* D_{\Omega,i,j}\left ( x \right ) \frac{dE_j}{dx} \right |_{x_r^-}^{x_r^+} + \left ( \int_{0}^{x_r^-} dx + \int_{x_r^+}^{1} dx \right ) \left \{ \frac{dE_i^*}{dx} \left (  \left (D_{\Omega,i,j} \left ( x \right ) - A_{\Omega,i,j} \left ( x \right )  \right )  \right ) \frac{dE_j}{dx} \right \} \right ] && \nonumber \\
\noindent = \left [ \left. E_i^* D_{k,i,j}\left ( x \right ) \frac{dE_j}{dx} \right |_{x_r^-}^{x_r^+} + \left ( \int_{0}^{x_r^-} dx + \int_{x_r^+}^{1} dx \right ) \left \{ E_i^* \left ( D_{k,i,j} \left ( x \right ) - A_{k,i,j} \left ( x \right ) \right ) E_j \right \} \right ] .
\end{eqnarray}
for all $E_m'$ continuous on $x \in \left (0 , x_r^- \right ) \cup \left (x_r^+ , 1 \right )$ where $E_m\left ( 0 \right ) = E_m'\left ( 0 \right ) = E_m\left ( 1 \right ) = E_m'\left ( 1 \right ) = 0$. Removing this part of the domain from the integration is equivalent to multiplying the integrand by a weight function $g\left (x \right )$ which is equal to $0$ for $x \in \left ( a,b \right )$ and $1$ elsewhere. The equation clearly lacks any information on the excluded region. Thus the equation expresses a necessary, but not sufficient condition for the solution $E_m$. Consequently, it is necessary to restrict $E_m$ in the excluded region to those solutions found using the Frobenius expansion. This can be achieved by replacing the finite element basis functions where $x \in \left [ x_r^-, x_r^+ \right ]$ with appropriate singular finite elements. Although, the above applies to cases with one continuum resonance, however it could readily be generalised to cases with multiple resonances.

\begin{figure}[h] 
\centering 
\includegraphics[width=80mm]{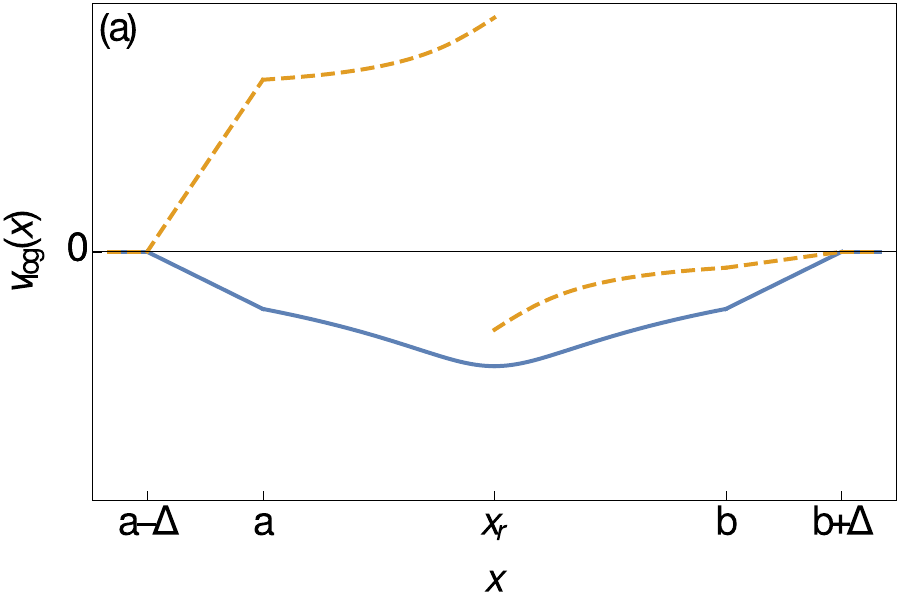} 
\includegraphics[width=80mm]{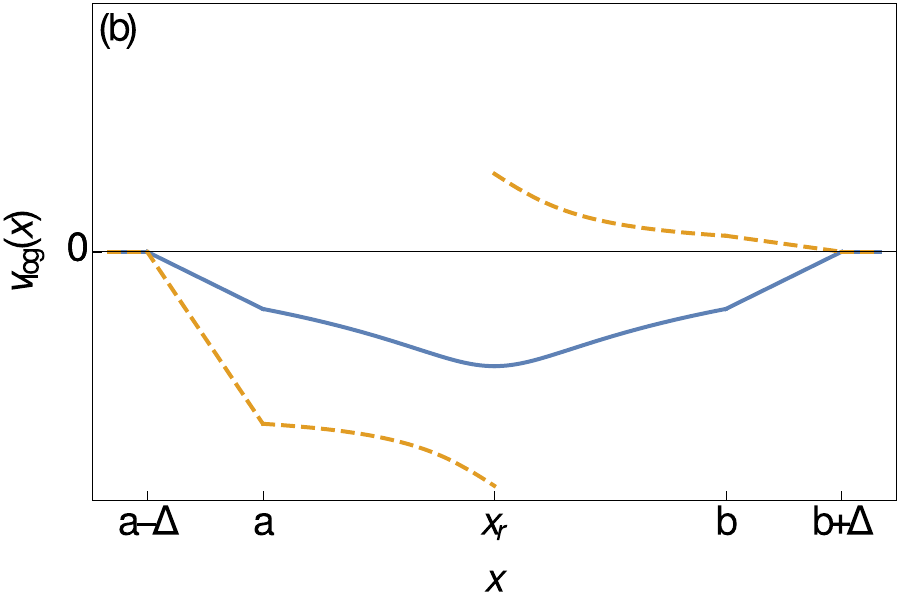} 
\caption{\label{fig:log_sfe} 
Plots (a) and (b) illustrate basis functions which are logarithmic on the inner region $x \in \left ( a,b \right)$, $v_{m,logarithmic,j}$ where $\Im \left ( x_r \right )$ is positive and negative respectively. In each case $\Im \left ( x_r \right )$ is finite, resulting in a continuous real component (blue, solid) and an imaginary component with a step-discontinuity at $x = \Re \left ( x_r \right ) $ (gold, dashed).
} 
\end{figure}

The basis functions described above are replaced with alternative functions reflecting the lowest order terms of the Frobenius expansion over an inner region $x \in \left (a , b \right )$. The bounds $a$ and $b$ are chosen such that $\left ( x_r^- , x_r^+ \right ) \subset \left (a , b \right )$ and both $a$ and $b$ are integer multiples of $\Delta$. The singular basis functions used are defined in terms of the analytically continued logarithmic function expressed in equation (\ref{eq:aclog}),
\begin{equation}
v_{m,log,j} = \left \{ \begin{array}{rcl}
\ln _{\pm} \left ( x - x_r \right ) \delta _{m,j} &\mbox{ for } x \in \left (a , b \right ) \\ 
\left (1 - \frac{\left (a-x \right )}{\Delta} \right ) \ln _{\pm} \left ( a - x_r \right ) \delta _{m,j} &\mbox{ for } x \in \left (a - \Delta , a \right ) \\
\left (1 - \frac{\left (x-b \right )}{\Delta} \right ) \ln _{\pm} \left ( b - x_r \right ) \delta _{m,j} &\mbox{ for } x \in \left (b , b + \Delta \right ) .
\end{array} \right.
\end{equation}
%\begin{equation}
%v_{m,log,j} = \begin{cases}
%\ln _{\pm} \left ( x - x_r \right ) \delta _{m,j} &\mbox{ for } x \in \left (a , b \right ) \\ 
%\left (1 - \frac{\left (a-x \right )}{\Delta} \right ) \ln _{\pm} \left ( a - x_r \right ) \delta _{m,j} &\mbox{ for } x \in \left (a - \Delta , a \right ) \\
%\left (1 - \frac{\left (x-b \right )}{\Delta} \right ) \ln _{\pm} \left ( b - x_r \right ) \delta _{m,j} &\mbox{ for } x \in \left (b , b + \Delta \right ) .
%\end{cases}
%\end{equation}
Such basis functions are illustrated in figure \ref{fig:log_sfe}. By including the discontinuity at $x = \Re \left ( x_r \right )$ due to the continuum resonance pole, such singular basis functions ensure that continuum damping is represented by the imaginary component of the eigenvalue.

\begin{figure}[h] 
\centering 
\includegraphics[width=80mm]{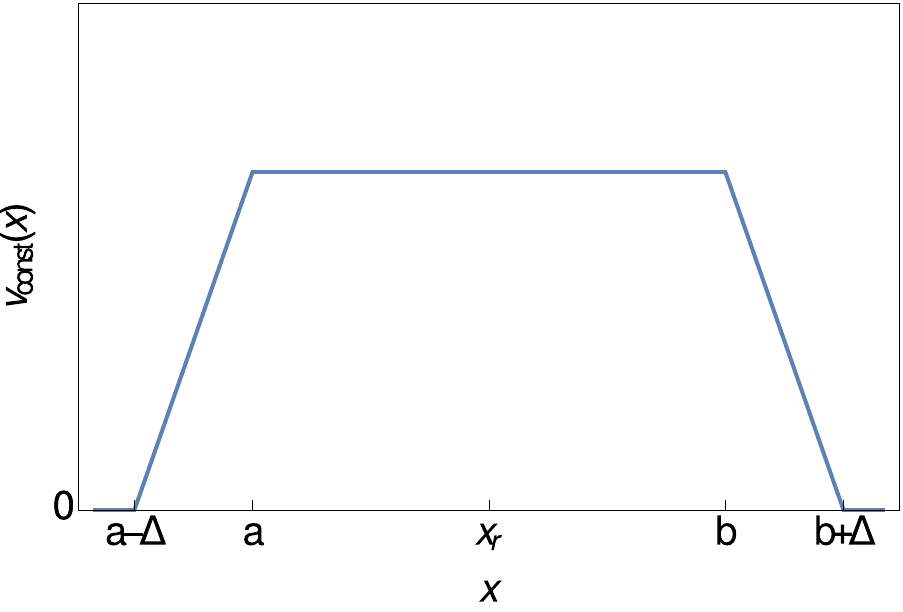} 
\caption{\label{fig:constant_fe} 
A plot illustrating a basis function which is linear over the inner region $x \in \left ( a,b \right)$, $v_{m,constant,j}$.
} 
\end{figure}

\begin{figure}[h] 
\centering 
\includegraphics[width=80mm]{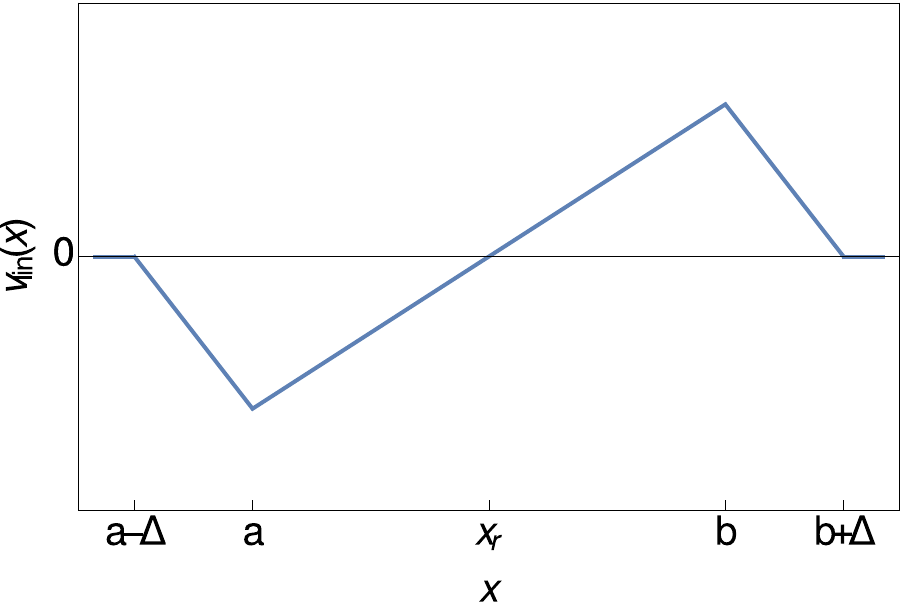} 
\caption{\label{fig:linear_fe} 
A plot illustrating a basis function which is linear over the inner region $x \in \left ( a,b \right)$, $v_{m,linear,j}$.
} 
\end{figure}

Elements which are constant on the inner region are also chosen, reflecting the constant terms in the expansion,
\begin{equation}
v_{m,const,j} = \left \{ \begin{array}{rcl}
\delta _{m,j} \textup{ for } x \in \left (a , b \right ) \\ 
\left (1 - \frac{\left (a-x \right )}{\Delta} \right ) \delta _{m,j} &\mbox{ for } x \in \left (a - \Delta , a \right ) \\
\left (1 - \frac{\left (x-b \right )}{\Delta} \right ) \delta _{m,j} &\mbox{ for } x \in \left (b , b + \Delta \right ) .
\end{array} \right.
\end{equation}
%\begin{equation}
%v_{m,const,j} = \begin{cases}
%\delta _{m,j} \textup{ for } x \in \left (a , b \right ) \\ 
%\left (1 - \frac{\left (a-x \right )}{\Delta} \right ) \delta _{m,j} &\mbox{ for } x \in \left (a - \Delta , a \right ) \\
%\left (1 - \frac{\left (x-b \right )}{\Delta} \right ) \delta _{m,j} &\mbox{ for } x \in \left (b , b + \Delta \right ) .
%\end{cases}
%\end{equation}
This type of basis function is illustrated in figure \ref{fig:constant_fe}. To improve convergence, basis functions were also defined which were linear on the inner region, representing the next lowest order terms in the expansion,
\begin{equation}
v_{m,lin,j} = \left \{ \begin{array}{rcl}
\left ( x - x_r \right ) \delta _{m,j} &\mbox{ for } x \in \left (a , b \right ) \\ 
\left (1 - \frac{\left (a-x \right )}{\Delta} \right ) \left (a - x_r \right ) \delta _{m,j} &\mbox{ for } x \in \left (a - \Delta , a \right ) \\
\left (1 - \frac{\left (x-b \right )}{\Delta} \right ) \left (b - x_r \right ) \delta _{m,j} &\mbox{ for } x \in \left (b , b + \Delta \right ) .
\end{array} \right.
\end{equation}
%\begin{equation}
%v_{m,lin,j} = \begin{cases}
%\left ( x - x_r \right ) \delta _{m,j} &\mbox{ for } x \in \left (a , b \right ) \\ 
%\left (1 - \frac{\left (a-x \right )}{\Delta} \right ) \left (a - x_r \right ) \delta _{m,j} &\mbox{ for } x \in \left (a - \Delta , a \right ) \\
%\left (1 - \frac{\left (x-b \right )}{\Delta} \right ) \left (b - x_r \right ) \delta _{m,j} &\mbox{ for } x \in \left (b , b + \Delta \right ) .
%\end{cases}
%\end{equation}
An illustration of such a basis function is provided in \ref{fig:linear_fe}. Inclusion of linear terms reflects the existence of a real component of the solution on the inner region which is anti-symmetric in both real and imaginary parts about the continuum resonance location.
% Last line is questionable, could be better worded

The location of the pole due to the continuum resonance is not known \textit{a priori}, as solutions to $\left \| D_{i,j} \right \| = 0$ are dependent on $\Omega$. If there is an error in the estimated resonance location, the basis functions above will not be able to accurately represent the solution near the resonance. Thus, the solution obtained will depend on the width of the excised region. Consequently it is necessary to apply the singular finite element using an iterative technique. The technique is applied as follows:
\begin{enumerate}
\item Compute the real frequency component using the finite element technique incorporating only standard linear elements. This is reasonably accurate, provided that the continuum damping is small in relation to the real frequency component.
\item Estimate the pole location $x_r$, by solving $\Omega_C \left ( x_r \right ) = \Omega$, where $\Omega _C$ is the normalised continuum frequency.
\item Add a small imaginary component to the estimated pole location $x_r$, reflecting that $\Omega _i < 0$ as required by the causality condition. The sign of this component is estimated based on a truncated Taylor series expansion which implies $\Im \left ( x_r \right ) \approx \Omega _i \left. \frac{\partial \Omega _C}{\partial x} \right | _{x=\Re \left ( x_{r} \right )} ^{-1}$.
\item Using singular finite elements, find $\Omega$ as a function of the width of the excised region. A larger excised width reduces the effect of the error in the pole location, though the remaining logarithmic part should be sufficiently large that it accurately reflects the solution.
\item Update the estimate for complex $\Omega$ by estimating the limit for the $\Omega$ based on the largest value of the excise width.
\item Update the estimate for the pole location. Approximate the real component using $\Omega_C \left ( \Re \left ( x_r \right ) \right ) \approx \Re \left ( \Omega \right )$ and then approximate the imaginary component based on the truncated Taylor series in step (iii).
\item Repeat the previous three steps to determine increasingly accurate values for $\Omega$ and $x_r$, which numerical experiment shows converge to constant values. As this occurs the dependence of $\Omega$ on the width of the excised region is removed.
\item Demonstrate convergence with respect to the number of radial grid points, $N$ and the width of the inner region, $b-a$.
\end{enumerate}

The complex contour method was used to verify the results obtained using singular finite elements. In this technique the eigenvalue problem is solved over a complex contour which is deformed around the complex poles due to continuum resonances in accordance with the causality condition \cite{computational_approach_to_continuum_damping_in_3D_published}. The complex contour chosen is parameterised by $ x=t+i \alpha t \left ( 1-t \right ) $, where $t \in \left (0,1 \right )$. A similar Galerkin method can be used to find the complex eigenvalues in this case, using a basis set composed exclusively of triangular functions. These basis functions are defined along the complex contour in terms of the contour parameter $t$. For the chosen \TAE\ case the equilibrium parameters $q\left (r \right )$ and $N\left (r \right )$ are analytic on the region of interest, allowing evaluation along the chosen complex contour.

\section{Verification} \label{sec:verification}
\begin{figure}[h] 
\centering 
\includegraphics[width=80mm]{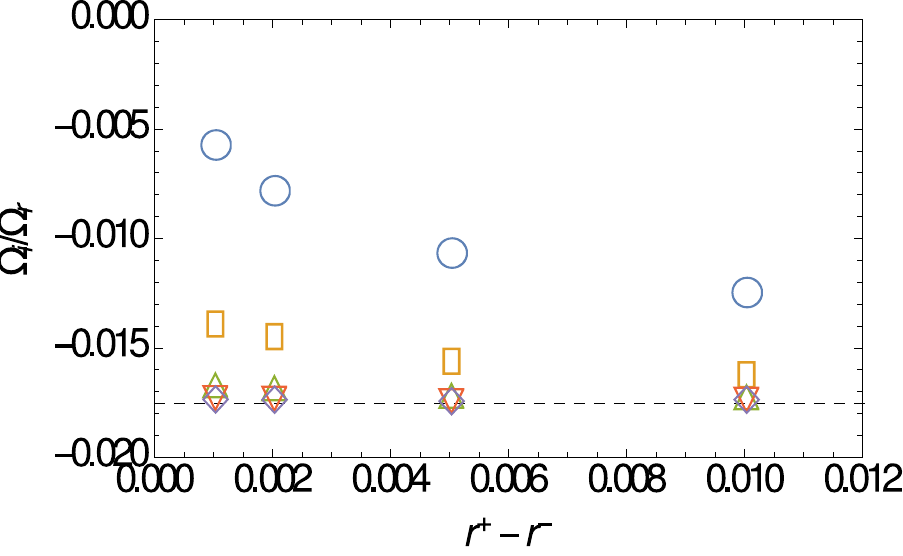} 
\caption{\label{fig:iterative_convergence} 
Damping ratio $\frac{\Omega_i}{\Omega_r}$ as a function of the width of the excised region $x_r^+ - x_r^-$ over five iterations using the singular finite element method. After each iteration $x_r$ is updated based on the value of $\Omega$ for $x_r^+ - x_r^- = 0.01$. The blue circle, gold box, green triangle, orange inverted triangle and purple diamond markers correspond to the first, second, third, fourth and fifth iterations respectively. The damping ratio found using the complex contour method is indicated by the black line.
} 
\end{figure}

\begin{figure}[h] 
\centering 
\includegraphics[width=80mm]{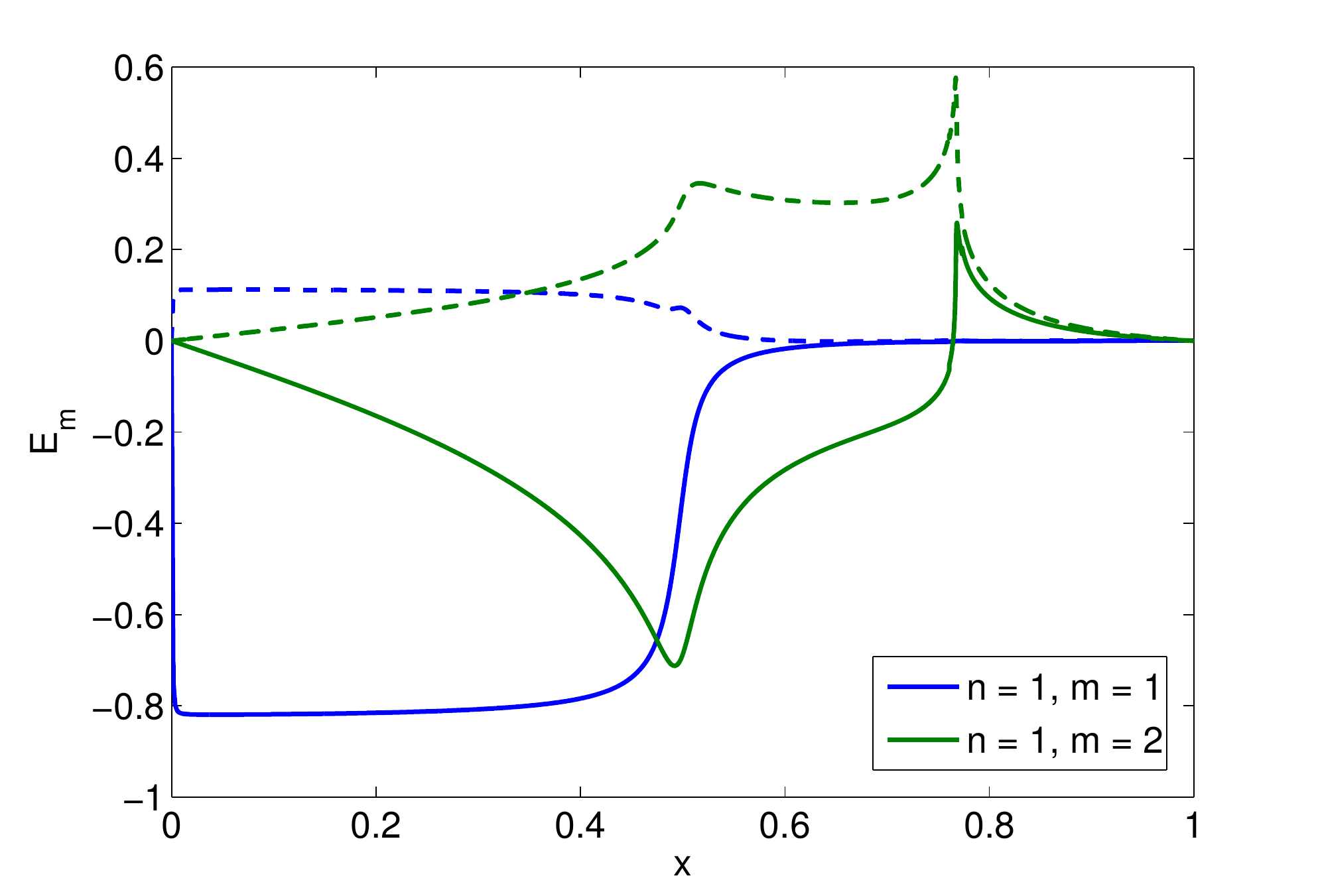} 
\caption{\label{fig:TAE} 
Mode structure for a \TAE\ due to coupling of the $\left ( m , n \right ) = \left ( 1 , 1 \right )$ and $\left ( 2 , 1 \right )$ harmonics with complex frequency $\Omega = 0.3258 -0.00569543 i$, found using the singular finite element method. Solid lines represent real quantities and dashed lines represent imaginary quantities. The continuum resonance pole is located at $x_r = 0.767504 -0.000511 i$. $N = 721$ radial grid points were used with inner region width $b-a = 0.0125$ and excised region width $x_r^+ - x_r^- = 0.005$.
} 
\end{figure}

\begin{figure}[h] 
\centering 
\includegraphics[width=80mm]{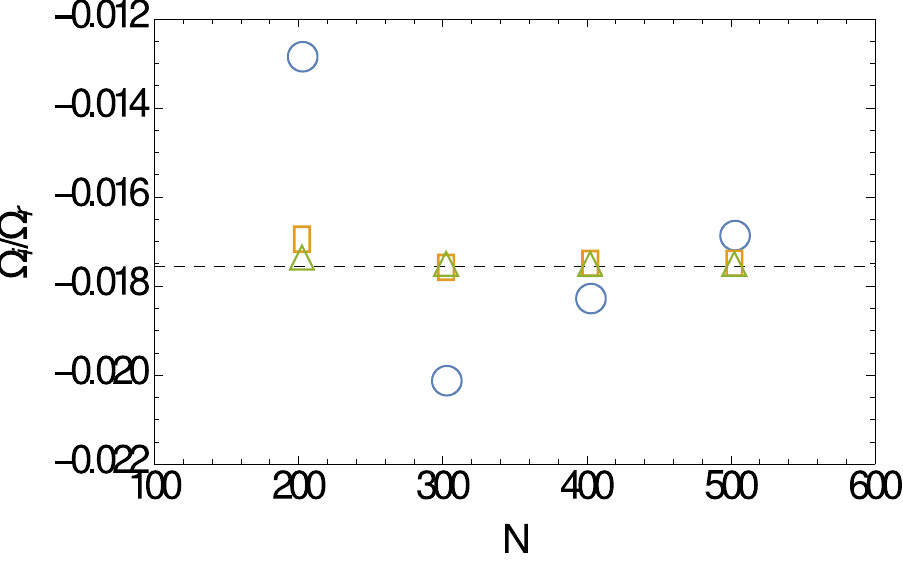} 
\caption{\label{fig:contour_convergence} 
Damping ratio as a function of the number of radial grid points $N$ for different contour deformation parameters $\alpha$, using the complex contour method. The blue circle, orange square and green triangle correspond to $\alpha = -0.01$, $-0.02$ and $-0.05$. The damping ratio found using the singular finite element method is indicated by the black line.
} 
\end{figure}

A \TAE\ mode due to the coupling of the $\left ( m , n \right ) = \left ( 1 , 1 \right )$ and $\left ( 2 , 1 \right )$ harmonics was studied using the simplified model outlined in section \ref{sec:singularities}. This analysis was done for a tokamak with aspect ratio $10$ ($\epsilon = 0.1$). The safety factor profile was chosen to be $q\left (x \right ) = q_0 + \left ( q_a - q_0 \right ) x^2$, where $q_0 = 1.0$ and $q_a = 3.0$. The plasma density profile was selected to be $N\left (x \right )=\frac{N_0}{2} \left (1-\tanh \left (\frac{x-\Delta_1}{\Delta_2}\right ) \right )$, where $\Delta_1 = 0.7$ and $\Delta_2 = 0.05$. A set of basis functions was chosen with the number of radial grid points $n = 401$ and inner region width $b - a = 0.0125$. The convergence of the damping ratio $\frac{\Omega _i}{\Omega _r}$ using the iterative technique outlined in section \ref{sec:FEM} is shown in figure \ref{fig:iterative_convergence}. After five iterations, it was estimated that the normalised frequency was $\Omega = 0.326 - 0.00572i$, corresponding to a damping ratio of $\frac{\Omega _i}{\Omega _r} = -0.0176$ (considering the case where the excised region had width $x_r^+ - x_r^- = 0.005$). A decrease in the inner region to $b-a = 0.0075$ and an increase to $b-a = 0.015$ altered the damping ratio by just $-0.81\%$ and $0.36\%$ respectively. Increasing the number of radial grid points used to $561$ and $721$ changed the damping ratio by just $-0.24\%$ and $0.03\%$ respectively, indicating convergence had also occurred with respect to this parameter. The mode structure obtained for the latter resolution is shown in figure \ref{fig:TAE}. Using a finite element complex contour method, the normalised frequency computed was $\Omega = 0.326 - 0.00571i$ and hence the damping ratio was $\frac{\Omega _i}{\Omega _r} = -0.0175$. The convergence of this result with respect to the deformation parameter $\alpha$ is shown in figure \ref{fig:contour_convergence}. The difference between the damping ratios computed using the contour and singular finite element techniques is $0.31\%$.

While the boundary condition $E_1 \left ( 0 \right ) = 0$ used here differs from that used by Berk \textit{et al.} \cite{continuum_damping_of_low_n_TAEs}, introducing this condition does not alter the results obtained to within three significant figures. In the singular finite element solution, $E_1$ assumes an approximately constant value very close to the magnetic axis as required for regular solutions of equation \ref{eq:wave_equation} near the origin. Using a shooting method implementation of the contour method, discussed in \cite{comparison_of_methods_for_numerical_calculation_of_continuum_damping}, with the boundary conditions of Berk \textit{et al.} gives $\frac{\Omega _i}{\Omega _r} = -0.0175$.

\section{Conclusion}
We have described a singular finite element method which successfully reproduces the \TAE\ frequency and continuum damping found using the complex contour method. As the continuum damping computed using the latter method has previously been shown to closely agree with the results of resistive \MHD\ \cite{computational_approach_to_continuum_damping_in_3D_published}, this agreement demonstrates the validity of the singular finite element method. The small errors in the results of these finite element methods are due to the limited accuracy inherent in approximating a solution with a limited number of finite elements. In the case of the singular finite element technique, these limitations arise due to the finite width of the singular and regular elements as well as the location of the pole used to construct the former.

The singular finite element technique presented here could be readily integrated into existing codes. This would be done by replacing standard finite elements with appropriate singular finite elements in the regions around continuum resonances. Unlike the complex contour technique, it does not require analytic continuation of equilibrium quantities. This is advantageous as finite element plasma stability codes typically employ numerical representations of these quantities which are based on spline interpolation and do not have analytic continuations over the domain of interest \cite{computational_approach_to_continuum_damping_in_3D_published}. In contrast to the complex contour method, the singular finite element method calculates the mode structure for real values of the radial coordinate $r$, rather than over a complex path in that variable. Moreover, less resolution is required to solve eigenvalue equations in ideal \MHD equations than resistive \MHD . Thus, the ideal \MHD\ singular finite element technique presented may allow calculation of continuum damping in more complicated geometries than has previously been practical, such as for stellarators. However, this singular finite element technique requires the user to demonstrate that the solution has converged with respect to four different parameters (grid resolution, excised width, singular element width and location). In contrast, convergence with respect to two parameters is required in the resistive technique (grid resolution and resistivity parameter) and the complex contour technique (grid resolution and contour deformation).

\section*{Acknowledgments}
We gratefully acknowledge the assistance provided by Em. Prof. R. L. Dewar, particularly in suggesting that a finite element approach be taken to the problem of continuum damping in plasma. We would also like to acknowledge helpful discussions regarding the continuum damping problem with Dr. G. R. Dennis.


\begin{thebibliography}{10}

\bibitem{damping_of_GAEs_in_tokamaks_due_to_resonant_absorption}
S.~Poedts, W.~Kerner, J.~P. Goedbloed, B.~Keegan, G.~T.~A. Huysmans, and
  E.~Schwarz.
\newblock {Damping of global Alfv{\'{e}}n waves in tokamaks due to resonant
  absorption}.
\newblock {\em Plasma Physics and Controlled Fusion}, 34(8):1397--1422, August
  1992.

\bibitem{landau_damping}
L.~Landau.
\newblock {On the vibrations of the electronic plasma}.
\newblock {\em JETP}, 16:574, 1946.

\bibitem{continuum_damping_of_high_n_TAWs}
M.~N. Rosenbluth, H.~L. Berk, J.~W. Van~Dam, and D.~M. Lindberg.
\newblock {Continuum Damping of High-Mode-Number Toroidal Alfv\'{e}n Waves}.
\newblock {\em Physical Review Letters}, 68(5):596--599, February 1992.

\bibitem{continuum_damping_of_ideal_TAEs}
X.~D. Zhang, Y.~Z. Zhang, and S.~M. Mahajan.
\newblock {Continuum damping of ideal toroidal Alfv\'{e}n eigenmodes}.
\newblock {\em Physics of Plasmas}, 1:381--389, February 1994.

\bibitem{resonant_damping_of_TAEs_in_tokamaks}
F.~Zonca and L.~Chen.
\newblock {Resonant Damping of Toroidicity-Induced Shear-Alfv\'{e}n Eigenmodes
  in Tokamaks}.
\newblock {\em Physical Review Letters}, 68:592--595, May 1992.

\bibitem{continuum_damping_of_low_n_TAEs}
H.~L. Berk, J.~W. Van~Dam, Z.~Guo, and D.~M. Lindberg.
\newblock {Continuum damping of low-n toroidicityinduced shear Alfv\'{e}n
  eigenmodes}.
\newblock {\em Physics of Fluids B}, 4:1806--1835, July 1992.

\bibitem{computational_approach_to_continuum_damping_in_3D_published}
A.~K{\"{o}}nies and R.~Kleiber.
\newblock {A computational approach to continuum damping of Alfv\'{e}n waves in
  two and three-dimensional geometry}.
\newblock {\em Physics of Plasmas}, 19, December 2012.

\bibitem{sfe_for_crack_propagation}
T.~Belytschko and H.~Chen.
\newblock Singular enrichment finite element method for elastodynamic crack
  propagation.
\newblock {\em International Journal of Computational Methods}, 1:2004, June
  2004.

\bibitem{edge_elements_for_eddy_current_problems}
{Oszk\'{a}r B\'{i}r\'{o}}.
\newblock Edge element formulations of eddy current problems.
\newblock {\em Computer Methods in Applied Mechanics and Engineering},
  169:391--405, February 1999.

\bibitem{sfe_for_Stokes_flow}
G.~C. Georgiou, L.~G. Olson, W.~W. Schultz, and S.~Sagan.
\newblock A singular finite element for stokes flow: The stick-slip problem.
\newblock {\em International Journal for Numerical Methods in Fluids},
  9:1353--1367, November 1989.

\bibitem{galerkin_method_for_DEs_with_regular_singular_points}
A.~D. Miller and R.~L. Dewar.
\newblock {Galerkin Method for Differential Equations with Regular Singular
  Points}.
\newblock {\em Journal of Computational Physics}, 66:356--390, October 1986.

\bibitem{Principles_of_MHD}
H.~Goedbloed and S.~Poedts.
\newblock {\em Principles of {M}agnetohydrodynamics}.
\newblock Cambridge University Press, Cambridge, UK, 2004.

\bibitem{comparison_of_methods_for_numerical_calculation_of_continuum_damping}
G.~W. Bowden, A.~K{\"o}nies, M.~J. Hole, N.~N. Gorelenkov, and G.~R. Dennis.
\newblock Comparison of methods for numerical calculation of continuum damping.
\newblock {\em Physics of Plasmas}, 21, 2014.

\end{thebibliography}
\end{document}